# Selective and Quasi-continuous Switching of Ferroelectric Chern Insulator Device for Neuromorphic Computing

Moyu Chen[1†], Yongqin Xie[1†], Bin Cheng[2*], Zaizheng Yang[1], Xin-Zhi Li[3], Fanqiang Chen[1], Qiao Li[1], Jiao Xie[1], Kenji Watanabe[4], Takashi Taniguchi[5], Wen-Yu He[3], Menghao Wu[6], Shi-Jun Liang[1*], Feng Miao[1*]

[1] Institute of Brain-Inspired Intelligence, National Laboratory of Solid State Microstructures, School of Physics, Collaborative Innovation Center of Advanced Microstructures, Nanjing University, Nanjing, China.

[2] Institute of Interdisciplinary Physical Sciences, School of Science, Nanjing University of Science and Technology, Nanjing 210094, China.

[3] School of Physical Science and Technology, ShanghaiTech University, Shanghai 201210, China

[4] Research Center for Functional Materials, National Institute for Materials Science, 1-1 Namiki, Tsukuba 305-0044, Japan.

[5] International Center for Materials Nanoarchitectonics, National Institute for Materials Science, 1-1 Namiki, Tsukuba 305-0044, Japan.

[6] School of Physics and School of Chemistry, Institute of Theoretical Chemistry, Huazhong University of Science and Technology, Wuhan 430074, China

† These authors contributed equally to this work.
* Correspondence Email:
bincheng@njust.edu.cn; sjliang@nju.edu.cn; miao@nju.edu.cn

**Topologically protected edge state transport in quantum materials is dissipationless and features quantized Hall conductance, and shows great potential in highly fault-tolerant computing technologies[1,2]. However, it remains elusive about how to develop topological edge state-based computing devices. Recently, exploration and**

**understanding of interfacial ferroelectricity in various van der Waals heterostructure material systems have received widespread attention among the community of materials science and condensed matter physics[3-11]. Such ferroelectric polarization emergent at the vdW interface can coexist with other quantum states[3,11] and thus provides an unprecedented opportunity to electrically switch the topological edge states of interest, which is of crucial significance to the fault-tolerant electronic device applications based on the topological edge states. Here, we report the selective and quasi-continuous ferroelectric switching of topological Chern insulator devices and demonstrate its promising application in noise-immune neuromorphic computing. We fabricate this ferroelectric Chern insulator device by encapsulating magic-angle twisted bilayer graphene with doubly-aligned h-BN layers, and observe the coexistence of the interfacial ferroelectricity and the topological Chern insulating states. This ferroelectricity exhibits an anisotropic dependence on the in-plane magnetic field. By using a $V_{BG}$ pulse with delicately controlled amplitude, we realize the nonvolatile switching between any pair of Chern insulating states and achieve 1280 distinguishable nonvolatile resistance levels on a single device. Furthermore, we demonstrate deterministic switching between two arbitrary levels among the record-high number of nonvolatile resistance levels. Such unique switching allows for the implementation of a convolutional neural network immune to external noise, in which the quantized Hall conductance levels of the Chern insulator device are used as weights. Our study provides a promising avenue toward development of the topological quantum neuromorphic computing, where functionality and performance can be drastically enhanced by topological quantum materials.**

## Main Text

The discovery of topologically protected edge states in quantum materials represents a milestone in the development of condensed matter physics. These topological edge states are not only featured with dissipationless transport but also characterized by different quantized Hall plateaus that persist within specific ranges of magnetic fields. Exploiting the edge states as information carriers allows for implementing novel computing electronic devices with highly fault-tolerant capability. Although this has been a long-desired goal of people working on topological quantum materials, it remains a huge challenge to realize the selective nonvolatile switching of the edge states by electric means, which is of crucial importance to electronic applications of these quantized edge states. Recently, the topologically protected edge states and interfacial ferroelectricity have been observed in the moiré heterostructures, respectively. By engineering the moiré heterostructures, it is feasible to achieve the coexistence of the interfacial ferroelectricity and the Chern insulating states in one single device. In this sense, the ferroelectric polarization emergent at the van der Waals interface can be used for switching the Chern insulating states, making the dissipationless device applications of the edge states possible.

In this work, we demonstrate selective and quasi-continuous switching in a ferroelectric Chern insulator device and its application in neuromorphic computing. The device was fabricated by aligning magic-angle twisted bilayer graphene (MATBG) with both top and bottom encapsulating h-BN crystals. Using electric pulses with delicately controlled amplitudes of gate voltage pulse, we realize selective and nonvolatile switching between any pair of Chern insulating states with distinct Chern numbers. By partially filling the Chern bands, we achieve a record-high number of 1280 distinguishable nonvolatile Hall resistance levels. Moreover, we demonstrate the deterministic switching between two arbitrary levels among these nonvolatile resistance levels. Based on such unique switching, we successfully train a convolutional neural network that uses the quantized Hall conductance as the weights and is immune to external noise. Our findings pave the way for topological quantum materials based neuromorphic computing.

**Observation of ferroelectricity in DA-MATBG device**

We fabricate dual-gated twisted bilayer graphene (TBG) samples encapsulated with h-BN flakes, with a schematic diagram and optical image shown in Fig. 1a and Supplementary Fig. S4, respectively, through the van der Waals assembly technique[12] (see Methods). Two monolayer graphene layers are stacked at the "magic angle" (~1.1°), where the electronic correlation can be remarkably enhanced to give rise to many exotic quantum states such as unconventional superconductivity[13-19] and correlated insulators[15,18,20-32]. Note that both top and bottom h-BN crystals are intentionally aligned with the MATBG sample. In this way, the rich moiré patterns at the graphene-BN and graphene-graphene interfaces could induce not only the electronic correlation, but also the spontaneous breaking of point group symmetries such as inversion and out-of-plane mirror reflection, which could facilitate the emergence of unconventional ferroelectricity in this bilayer graphene system[3,7,11]. To independently tune the out-of-plane displacement field $D_{ext}$ and carrier density $n_{ext}$ in the MATBG sample, we adopt few-layer graphite as top gate $V_{TG}$ and bottom gate $V_{BG}$, both of which are isolated from the MATBG sample by h-BN layers.

We perform electrical transport measurements on our MATBG device at 4K, with the corresponding results shown in Fig. 1b, c. With $V_{BG}$ grounded, the longitudinal resistance $R_{xx}$ as a function of $V_{TG}$ exhibits prominent resistance peaks at $V_{TG} = 0, \pm 3V$, corresponding to the charge neutrality point (CNP) and band insulators (BIs) with the moiré flat bands either empty or fully occupied of electrons with 4-fold spin-valley degeneracy. When the flat bands are partially filled, resistance peaks emerge at $V_{TG} = 0.8$ V and $V_{TG} = 1.6$ V. These two resistance peaks correspond to correlated states (CSs), with filling factors $\nu$ equal to 1 and 2, respectively. We observe no hysteresis of $R_{xx}$ in sweeping of $V_{TG}$, while robust hysteretic behavior appears when sweeping $V_{BG}$ forward and backward (Fig. 1c). It is worthwhile to point out that $R_{xx}$ curve exhibits no sudden jump as we sweep $V_{BG}$, which can be attributed to $V_{BG}$ induced continuous variation of ferroelectric polarization (see Supplementary Fig. S2, S3). To further evaluate the $V_{BG}$ dependence of ferroelectric polarization, we measure $R_{xx}$ as a function of the sweeping window $\Delta V_{BG}$, defined as $\Delta V_{BG} = V_{BG}^{max} - V_{BG}^{min}$, while keeping $V_{TG}$ grounded (see Supplementary Information Section I). By tracking the CNP where $n = 0$, we extract remnant polarization $2p_r$ (see Supplementary Information Section I) and present the results in Fig. 1d. Notably, the remanent polarization is not effectively activated until the sweeping window $\Delta V_{BG}$ reaches a threshold (i.e., 14 V), above which $2p_r$ linearly depends on $\Delta V_{BG}$.

The emergent ferroelectricity in our MATBG sample indicates spontaneous layer-polarized occupation of flat-band electrons[7]. We use the spontaneous polarization in the bilayer graphene system to estimate such layer-polarized electron occupation, denoted as $\Delta \nu \equiv (\nu_T - \nu_B)/2$ with $\nu_T, \nu_B$ the number of electrons per moiré unit cell occupying the top or bottom graphene layer (see Supplementary Information Section II). As shown in

Supplementary Fig. S3, the $\Delta\nu$ is linearly dependent on $V_{BG}$ until it saturates to $\sim 4$, suggesting that nearly all the strongly correlated electrons in the flat bands only occupy one of graphene layers in MATBG sample. Such a large ratio of electron occupation between two graphene layers gives rise to a remnant polarization $p_r$ ($\sim 0.67\ \mu C/cm^2$) which is comparable to those reported in other 2D ferroelectric systems, e.g. $\sim 0.02\ \mu C/cm^2$ in 1T' $WTe_2$[33], $\sim 0.74\ \mu C/cm^2$ in AB-stacked bilayer h-BN[9,10], and $\sim 0.32\ \mu C/cm^2$ in 3R-stacked TMDs[4,5].

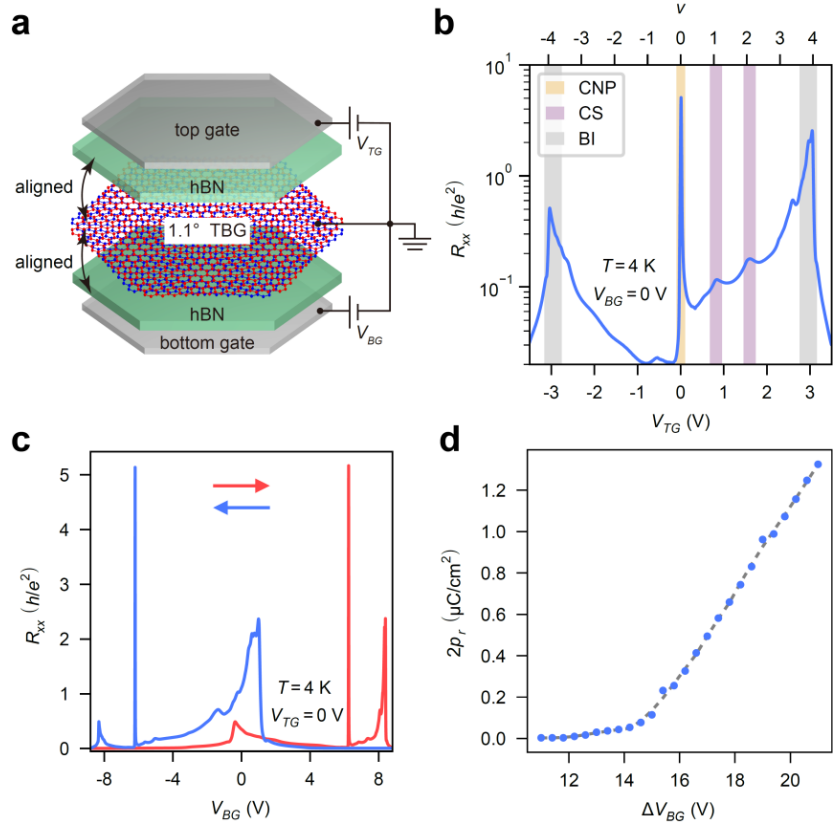

**Fig. 1. Characterization of DA-MATBG devices. a**, Schematic of DA-MATBG device structure with top ($V_{TG}$) and bottom ($V_{BG}$) gates. The graphene layers are double-aligned with the encapsulating h-BN crystals. **b**, Four-probe resistance $R_{xx}$ as a function of $V_{TG}$ at 4K, highlighting charge neutrality point (CNP), correlated states (CS) and band insulators (BI) features. **c**, Hysteresis in $R_{xx}$ as a function of $V_{BG}$ at 4K. The forward and backward scans are shown in red and blue, respectively. **d**, The remanent polarization $2p_r$ as a function of the $V_{BG}$ sweeping range, $\Delta V_{BG} = V_{BG}^{max} - V_{BG}^{min}$. The dashed line is a guide to the eye.

**Anisotropic dependence of ferroelectric polarization on in-plane magnetic field**

We next proceed to study the in-plane magnetic field dependence of the polarization in our DA-MATBG device. By slowly stepping $V_{BG}$ between $-10$V and $+10$V while sweeping $V_{TG}$ at each step of $V_{BG}$, we perform dual-gate mappings of $R_{xx}$ under zero and finite in-plane magnetic field ($B_\parallel = 12$ T). In Fig. 2a, b, we present the mapping of the difference of $R_{xx}$ between $V_{BG}$ stepping forward and backward, denoted as $\Delta R_{xx} = R_{xx}^{for} - R_{xx}^{back}$, exhibiting ferroelectric hysteresis as indicated by bow-shaped loops. These hysteresis loops at 12T (Fig. 2b) deviate from those under zero $B_\parallel$ (Fig. 2a), giving rise to an unconventional magnetic response of ferroelectricity previously unidentified in moiré heterostructures. When we reduce the in-plane magnetic field, the remnant polarization is decreased and exhibits a linear $B_\parallel$ dependence (Fig. 2c). By rotating the in-plane magnetic field, we observe an anisotropic azimuthal angle $\theta$ dependence of remnant polarization $2p_r$ (Fig. 2d), revealing that the magnetoelectric coupling has broken rotational symmetry. This magnetoelectric effect is independent of the current direction (see Supplementary Fig. S7), further indicating that the rotational symmetry breaking arises from the inherent origin e.g., formation of angle domains and heterostrain, both of which have been experimentally identified by the Raman spectroscopy[34], scanning tunneling microscopy[30,35-37], transmission electron microscopy[38,39]. Different from the in-plane field, applying an out-of-plane magnetic field $B_\perp$ to the device has no impact on $2p_r$ (see Supplementary Fig. S8). Because the polarization is along the out-of-plane direction and is thus orthogonal to the in-plane magnetic field, we term such effect as orthogonal magnetoelectric effect. Such magnetoelectric effect in the MATBG system is quite unusual since carbon, boron and nitrogen atoms are all nonmagnetic and contribute negligible spin-orbit coupling.

Through symmetry analysis, we can attribute the orthogonal linear magnetoelectric effect to the lowering of point group symmetry, which is ubiquitous in the moiré heterostructure due to the inevitable strain[38,39]. As seen in Supplementary Table S2, such coupling is allowed only when rotational symmetry is broken. By adopting a phenomenological model (i.e., generalized Landau-Khalatnikov model) including the magnetoelectric coupling term with the allowed form of linear magnetoelectric susceptibility pseudotensor $\alpha_{ij}$ (as indicated in Supplementary Table S2), we can immediately obtain an in-plane magnetic field induced shift in the ferroelectric hysteresis (see Supplementary Fig. S16), which can explain our experimental observations. Remarkably, our orthogonal magnetoelectric coupling effect exhibits a giant coupling parameter $\alpha \equiv \frac{dp_r}{dB_\parallel} \approx \frac{3e^2}{h}$, which is among the highest values reported in all reported literatures.

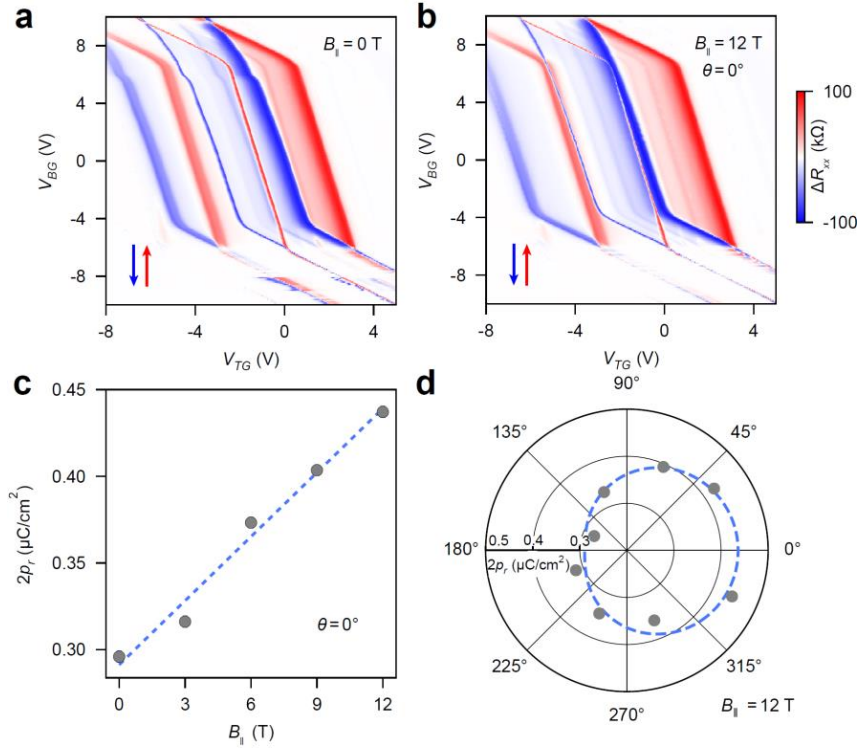

**Fig. 2. In-plane magnetic field dependence of remanent polarization. a**, Dual-gate mapping of four-probe resistance difference $\Delta R_{xx} = R_{xx}^{for} - R_{xx}^{back}$, with $R_{xx}^{for}$ ($R_{xx}^{back}$) denoting $R_{xx}$ with $V_{BG}$ stepping forward (backward). The presence of hysteresis loops indicates the emergent ferroelectricity. **b**, Same as **a** under $B_\parallel = 12T$ along azimuthal angle $\theta = 0°$. The hysteretic loop substantially deviates from that in **a**. **c**, Extracted remnant polarization $2p_r$ as a function of in-plane magnetic field $B_\parallel$ along $\theta = 0°$. The blue dashed line is a guide to the eyes. **d**, Remnant polarization $2p_r$ as a function of $\theta$ while keeping the magnitude of in-plane magnetic field $B_\parallel = 12T$. The blue dashed line is a guide to the eyes. Here, $\theta = 0°$ is set to the orientation of $B_\parallel$ where the largest value of $2p_r$ is obtained. Data in **c**, **d** are based on Supplementary Fig. S5 and S6.

**The selective and quasi-continuous switching in ferroelectric Chern insulators**

We show that the nontrivial topology of the strongly correlated flat bands in the MATBG can be imprinted into the ferroelectricity to give rise to nonvolatile topological states. Note that the Chern number has been widely used to characterize the topology of electronic band structures in graphene moiré systems. Nevertheless, the flat subbands in different valleys and of opposite Chern numbers are degenerate. To identify nonvolatile topological states imprinted with nontrivial topology, we apply an out-of-plane magnetic field $B_\perp$ to rearrange those topological subbands[29] and measure the longitudinal resistance $R_{xx}$ and transverse resistance $R_{xy}$. Figure 3a and 3b shows $R_{xx}$ and $R_{xy}$ as a function of $V_{TG}$ and $B_\perp$, respectively. A series of trajectories that exhibit $R_{xx}$ minima and $R_{xy}$ plateaus are identified under $B_\perp > 8$ T. These trajectories follow the Diophantine equation $\nu = C n_\phi + s$, where $s, C \in \mathbb{Z}$ correspond to the superlattice filling factor under zero magnetic field and the total Chern number of the occupied energy bands, respectively, and $n_\phi = \phi/\phi_0$ represents the number of quantum flux per superlattice unit cell[40]. Based on these trajectories, a series of energy levels, labelled by $(s, C) = (4, 0), (3, 1), (2, 2)$ and $(1, 3)$, can be identified, manifesting the successive integer filling of the topological nontrivial moiré subbands. Notably, this successive filling of the subbands with nonzero Chern numbers suggests that the flavor (spin/valley) degeneracy of the subbands is fully lifted, as a result of the strong e-e interaction and the presence of $B_\perp$ [24,29]. By coding the band topology into the emergent ferroelectricity, we thus realize ferroelectric Chern insulators in our DA-MATBG, in which the layer, spin and valley degrees of freedom are all polarized and intertwined. Such coexistence of ferroelectricity and Chern insulating states offers an unprecedented opportunity to switch the Chern bands that have spontaneously polarized spin and valley by tuning the spontaneously polarized layer degree of freedom corresponding to the ferroelectric polarization.

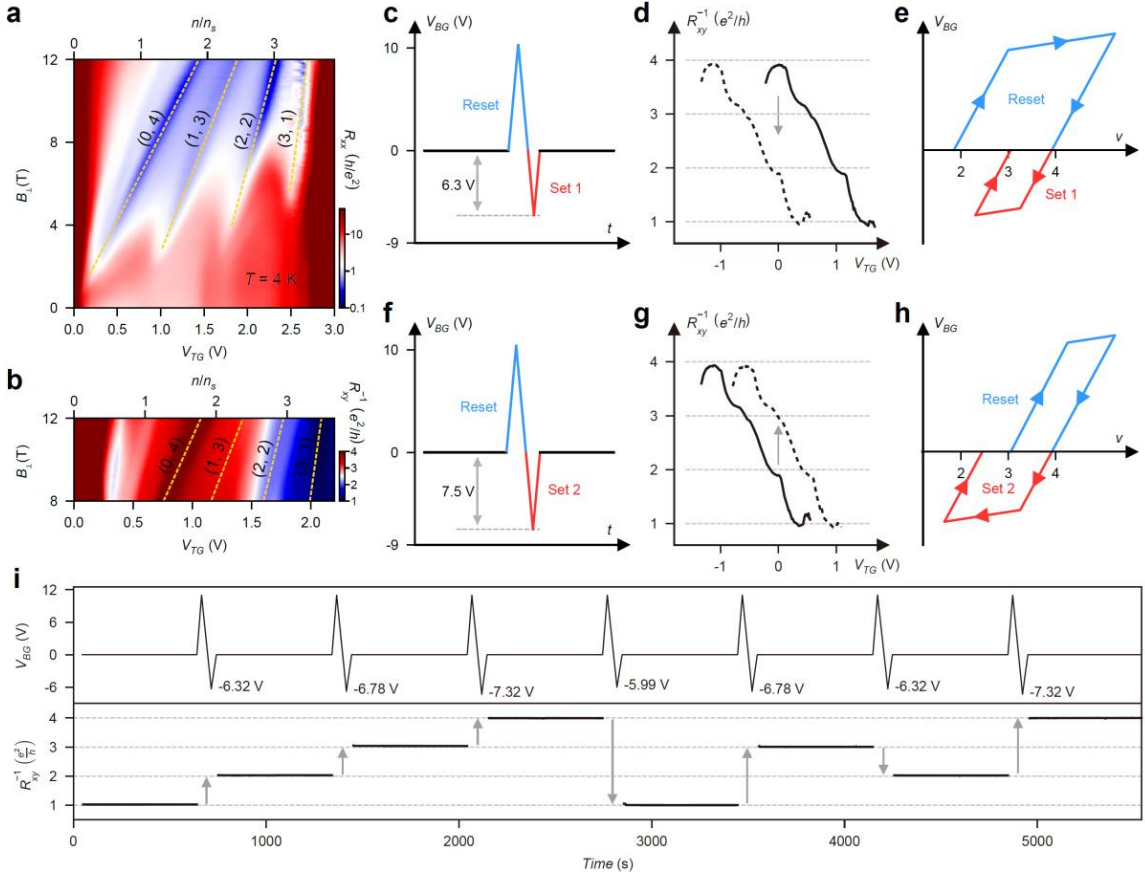

**Fig. 3. Ferroelectric Chern insulators and selective switching. a**, Longitudinal resistance $R_{xx}$ as a function of $V_{TG}$ and $B_\perp$. The yellow dashed lines indicate the trajectories of four Chern insulating states with $(s, C)$ indices of $(0,4), (1,3), (2,2)$ and $(3,1)$. **b**, Inverse transverse resistance $R_{xy}^{-1}$ as a function of $V_{TG}$ and $B_\perp$, exhibiting quantized plateaus of the Chern insulators. **c**, Waveform of $V_{BG}$ pulse to realize selective switching between Chern insulating states. Each waveform is composed of a reset pulse (blue) and a set pulse (red). **d**, $R_{xy}^{-1}$ as functions of $V_{TG}$ before (dashed line) and after (solid line) $V_{BG}$ pulse, representing a nonvolatile switching from Chern insulating state of $(0,4)$ to $(2,2)$. $V_{BG}$ is kept grounded as $V_{TG}$ sweeps. **e**, Schematic of the pulse-controlled open trajectory in the plane of the filling factor $\nu$ and $V_{BG}$. Blue and red line segments correspond to reset and set process, respectively. **f-h**, Demonstration of another nonvolatile switching from the Chern insulating state with $(2,2)$ to that with $(1,3)$ by applying an electric pulse of different amplitude. **i**, Selective switching among distinct Chern insulating states with a series of $V_{BG}$ pulses. The applied $V_{BG}$ waveform is shown in the upper panel

with the amplitudes of negative parts annotated near each pulse, while the measured $R_{xy}$ is shown in the lower panel.

We then demonstrate the nonvolatile switching of these Chern insulating states by using a special $V_{BG}$ double pulses including positive and negative, as shown in Fig. 3c, to change the ferroelectric polarization. Amplitudes of the positive pulse are kept at 10.4 V in the following experiments so that the device is reset to the fully polarized state before applying the negative part to set to the target Chern insulating state. Before and after applying each $V_{BG}$ pulse, $V_{TG}$ is swept to monitor $R_{xy}$, so that the initial and final Chern insulating states can be identified. For example, we can use an electric pulse as shown in Fig. 3c to switch the Chern insulating state with $C = 4$ to that with $C = 2$, which corresponds to a substantial shift of the $R_{xy}$ curve while its profile remains unchanged (Fig. 3d). This nonvolatile switching is enabled by the linear dependence of remnant polarization $2p_r$ on $\Delta V_{BG}$, which leads to an open trajectory with selected starting and ending points in the $\nu$-$V_{BG}$ phase plane (Fig. 3e) given the applied $V_{BG}$ pulse. Since the positive part of the pulse resets the polarization and $\nu$, the final state only depends on the amplitude of the negative part. In this way, we can achieve selective nonvolatile switching between any pair of Chern insulating states by delicately controlling the amplitude of the pulse (see Supplementary Information Section VI). Figure 3f-h illustrate another example of switching from the Chern insulating state with $C = 2$ to that with $C = 3$. We further demonstrate that the selective switching can be extended to other nonvolatile Chern insulating levels, as labelled by the arrows in Fig. 3i. Note that the selective switching of the topological Chern insulating states has not yet been reported before[26,41,42]. Although electric control of Chern insulating states has been reported in graphene-based magnetic Chern insulators[26,42], the Chern insulating states have to be retained by the non-zero gate voltage. In contrast, our ferroelectric Chern insulators can be maintained even if the gate voltage is removed, which is critical to designing the topological memory device with ultralow power consumption.

By partially filling Chern bands, we can realize a plethora of nonvolatile Hall resistance levels. As shown in Fig. 4a, we demonstrate nonvolatile switching among 1280 Hall resistance levels interspersed in the intervals of the quantized Chern insulating plateaus ranging from $\frac{h}{4e^2}$ to $\frac{h}{e^2}$, by applying a series of $V_{BG}$ pulses (see Supplementary Information Section VI). In Fig. 4b, we zoom in and display non-overlapping and nearly evenly spaced 100 resistance states with an average interval of ~$0.0005\frac{h}{e^2}$. Those small intervals between adjacent nonvolatile resistive levels correspond to extremely small changes of ferroelectric polarization on the order of $10^{-5} \mu C/cm^2$, which is about three orders of magnitude smaller than that accessible in the conventional ferroelectric materials (see Fig. 4c). The change of total electric dipole can be characterized by $\delta p_r \cdot V = e \delta n_p \cdot d \cdot S$, where $\delta p_r, V, e, \delta n_p, S, d$ denote the change of polarization density, volume, elementary charge, variation of layer-polarized charge density, area and thickness of DA-MATBG, respectively. With $V = S \cdot d$, $\delta p_r$ can be associated with $\delta n_p$ through the formula of $\delta n_p = \delta p_r / e$. We show that the variation of layer-polarized charge density $\delta n_p$ can be as small as $4 \, \mu m^{-2}$ (Fig. 4c, see Supplementary Information Section II for details), indicating that the resolution of ferroelectric switching can approach elementary charge limit in a submicron-sized device. Such subtle change in the layer-polarized charge density using $V_{BG}$ pulsing leads to the quasi-continuous charge polarization in the ferroelectric Chern insulators (Fig. 4d).

This quasi-continuous switching of these nonvolatile Hall levels can be attributed to the spontaneous layer-polarized occupation of topologically nontrivial electrons in the Chern bands. As illustrated in Fig. 4d, bottom-layer-polarized electrons can be injected into the Chern insulator by applying a $V_{BG}$ pulse. The electron filling is then stabilized with an increase in spontaneous polarization denoted by $\delta p_r = e \delta n_p$ after $V_{BG}$ is turned back and grounded, with $\delta n_p$ the change of layer-polarized charge density and $e$ the electron charge. The injected electrons and the corresponding change of ferroelectric polarization can lead to the change of transverse conductance through the formula $\delta R_{xy}^{-1} = -\frac{2e^2}{h}\Omega \delta n_p = -\frac{2e}{h}\Omega \delta p_r$, where $\Omega$ is the Berry curvature of the Chern bands. Therefore, the injected electrons carry both electric dipole and Berry curvature, thus simultaneously contributing to the increase of spontaneous polarization and Hall resistance in our ferroelectric Chern insulators. The merit of the ferroelectric Chern insulator device is not limited to the selective and quasi-continuous switching. Notably, the Chern insulating states are characterized by the quantized plateaus of the Hall resistance, which maintain their values despite variations in external control parameters, such as electric or magnetic fields, within a specific range. Thereby, it is promising to use these quantized conductance values of the device as the weights of the convolutional neural network to realize neuromorphic computing that can be immune to external noise, which has been a grand challenge in the field of neuromorphic computing[43-47].

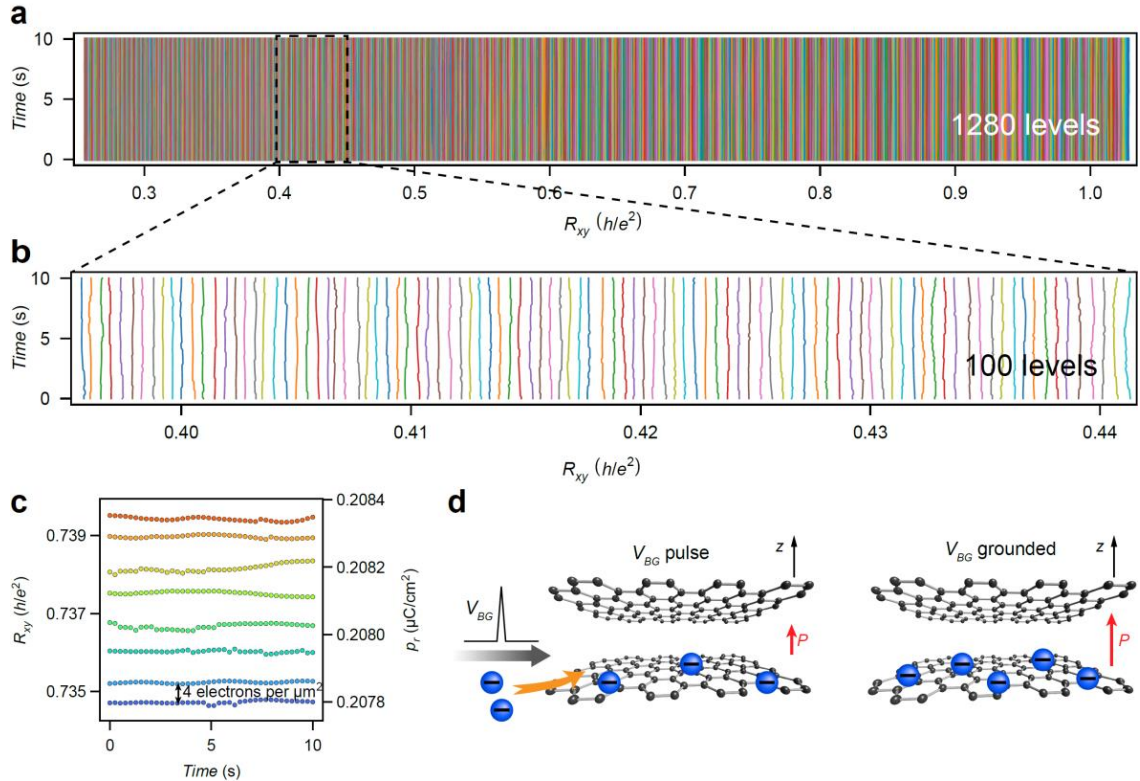

**Fig. 4. Quasi-continuous ferroelectric switching. a,** Transverse resistance $R_{xy}$ is switched to 1280 nonvolatile states by dynamically controlled $V_{BG}$ pulses. **b,** Zoomed-in view of 100 states from $0.396h/e^2$ to $0.442h/e^2$, showing that all the states are non-overlapping and nearly evenly spaced. **c,** A set of nearest quasi-continuous nonvolatile states from data in **a** and their corresponding polarization $p_r$. The polarization is calculated using the change of $V_{TG}$ that is required to produce the same change in $R_{xy}$ (see Supplementary Information Section II). The arrow denotes the corresponding change in $R_{xy}$ and $p_r$ between the nearest resistance levels, representing that the density of layer-polarized electrons changes by ~4 electrons per square micron. **d,** Schematic of nonvolatile switching of quasi-continuous polarization levels. A $V_{BG}$ pulse injects electrons into the bilayer graphene system, as shown in the left panel. The bottom-layer-polarized electron is retained when $V_{BG}$ is turned off and grounded, as shown in the right panel.

**Deterministic switching between two arbitrary levels among quasi-continuous levels**

Reproducibly accessing these quasi-continuous distinct levels on demand is key for neuromorphic computing application. By developing a generic physical mode used for delicately controlling the electric pulse, we are able to reproducibly access arbitrarily selected resistance levels among these quasi-continuous resistance levels (see Supplementary Fig. S13 and Supplementary Information Section VII). For demonstration of deterministic switching between two arbitrary levels among quasi-continuous levels, we label these distinct resistance levels with three-digit numbers, *i.e.,* from "000" to "999", as shown in Fig. 5a. To choose these levels, we use each three digits of the value of π to represent a three-digit number, and use the first three digits "3, 1, 4" of the value of π to represent the initial levels coded with "314" (Fig. 5b). Based on the switching protocol displayed (see details in Supplementary Fig. S13), we experimentally achieve the deterministic switching of two arbitrary levels, with the results shown in Fig. 5c. Distinct target resistance levels are represented with grey dotted lines in the lower panel of Fig. 5c. We find that applying 2 to 5 pulses (see top panel of Fig. 5c) allows for successful access of the target resistance levels within a tolerable deviation. Remarkably, this developed switching protocol yields high-precision switching without the need to reset the whole system, and can be readily adopted to realize the switching between arbitrary resistance levels available in the ferroelectric Chern insulator device.

Such deterministic switching allows us to train a convolutional neural network, in which the dissipationless chiral edge states with distinct quantized Hall resistance are used as weights. The trained weights of the convolutional neural network are then quantized with the quantized Hall conductance available in our ferroelectric Chern insulator device, as shown in Fig. 5d, 5e. To show the advantage of using the topologically protected Chern insulator states in the neuromorphic computing, we performed comparison of the inference accuracy of convolutional neural network between our device and traditional ReRAM device, with the results shown in Fig. 5f. The recognition accuracy of the neural network using the Chern insulator states is independent of the external noise, while the noise has a profound impact on the performance of the neural network based on the ReRAM device. This result strongly indicates that the significance of topological protection of the Chern insulator states in the neuromorphic computing, which is inaccessible with other previously reported technologies.

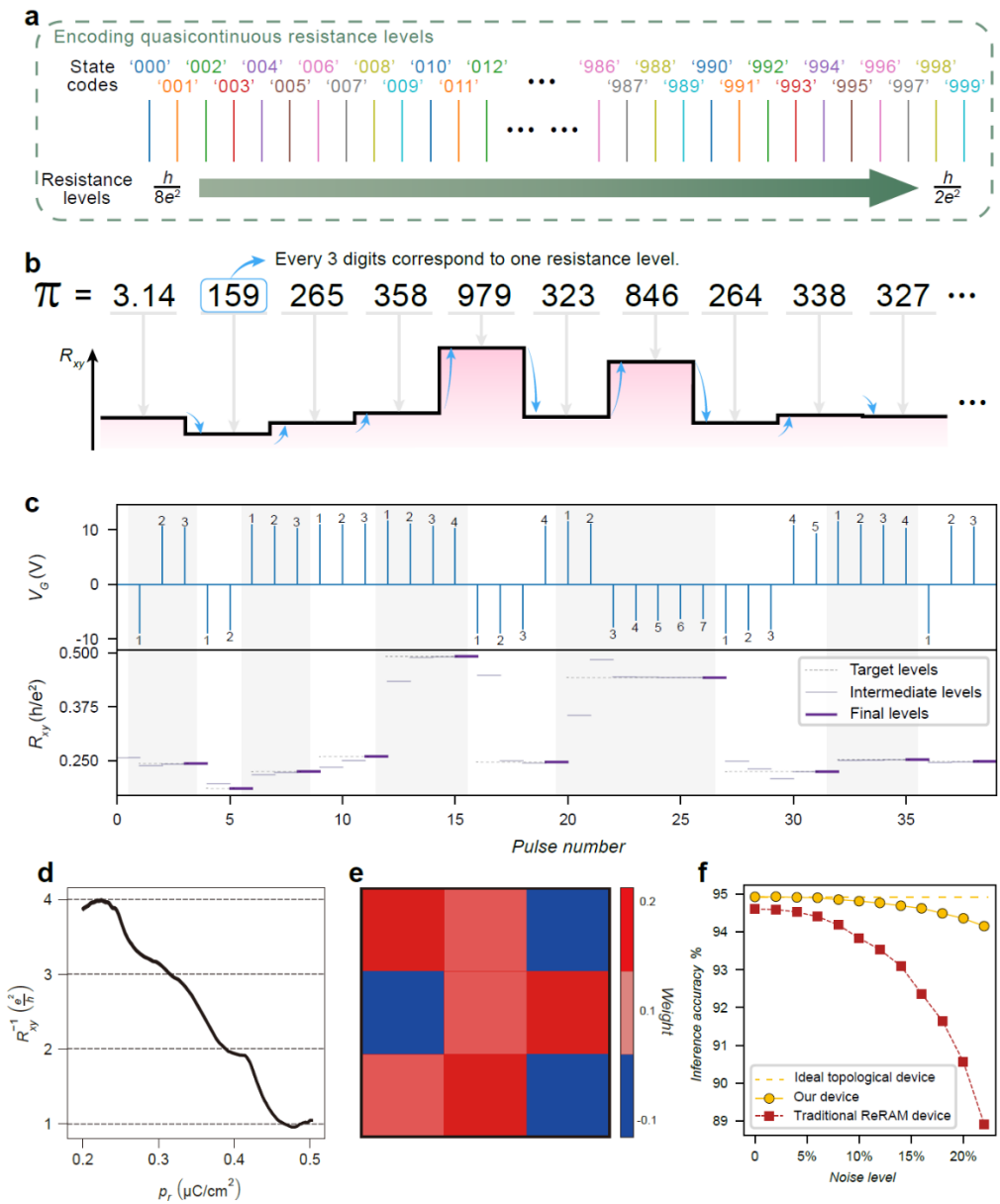

**Fig. 5. Selective switching between quasi-continuous resistance levels. a**. A schematic illustration of the correspondence between state codes and resistance levels. The three-digit state codes are labelled from '000' to '999' to represent the 1000 quasi-continuous resistance levels from $h/8e^2$ to $h/2e^2$. These 1000 resistance levels exhibit an average separation of 9.6Ω. **b**. The schematic of a series of selective nonvolatile switching among 10 random resistance levels, which are determined in the following demonstration: We start with the first 3 consecutive digits in $\pi$ as the first state code '314', then move on to the next 3 digits for a subsequent state code and resistance level and so on. **c**. Experimental results of selective switching. For each target resistance level discriminated by different background colors, we employ a few consecutive pulses where the pulse indices are denoted by the numbers over each pulse, as shown in the top panel. The bottom panel shows the stabilized resistance level after each pulse. The intermediate resistance levels are shown by pale purple lines while the final levels are shown by purple lines. The target levels are denoted by gray dashed lines. **d.** The transverse conductance $R_{xy}^{-1}$ as a function of the remnant polarization $p_r$. $R_{xy}^{-1}$ is linearly mapped to the weights shown on the right y-axis. The dashed lines denote the quantized conductance and weights. **e.** An example of the trained kernel. All trained weights are distributed on the quantized levels. **f.** Simulated inference accuracy of the convolutional neural network as a function of noise level for our device and traditional ReRAM device, respectively. The yellow dashed line is a guide to the eye representing the accuracy performance of ideal topological device.

**Conclusion and Outlook**

In summary, we observe the coexistence of vdW interfacial ferroelectricity and topological edge states in ferroelectric Chern insulators device based on DA-MATBG material. This ferroelectricity exhibits an anisotropic dependence on the in-plane magnetic field, giving the first example of orthogonal magnetoelectric effect in the family of moiré heterostructures. Moreover, we demonstrate the selective and deterministic switching of the Chern insulating states and quasi-continuous Hall resistance levels with Chern bands partially filled. Based on the unique switching, we employ the Chern insulating state as the weights to demonstrate a noise-tolerant convolutional neural network. Although, as a proof-of-concept, our device requires an external magnetic field to break time-reversal symmetry, the control methods and noise-tolerant neural network could be generalized to other Chern insulator systems including zero-field Chern insulators. Our work not only provides a new tool for nonvolatile tuning of topological quantum states by electric means, but also opens up an avenue for neuromorphic computing based on topological quantum materials.

# Methods

## Sample fabrication

We fabricated the DA-MATBG devices by a twist-angle-controlled dry-transfer method[12]. We first mechanically exfoliated monolayer graphene, few-layer graphite and h-BN crystals on $SiO_2$/Si substrates, and carefully selected flakes of appropriate thickness and high quality by optical contrast and atomic force microscopy (AFM). To avoid strain-related effects, we cut the monolayer graphene into two pieces with an AFM tip using the electrode-free anodic oxidation method[18] prior to the transfer process. The heterostructure assembly started with picking up a few-layer graphite flake and a h-BN flake sequentially using a poly (bisphenol A carbonate) (PC)/ polydimethylsiloxane (PDMS) stamp at 80°C. We then picked up the first piece of graphene flake at room temperature, followed by rotating the second piece by 1.1° and picking up the second piece. At last, we encapsulated the twisted bilayer graphene with another h-BN flake and a few-layer graphite flake at 80°C, before releasing the heterostructure from the PC film to a $SiO_2$/Si substrate at 140°C. Notably, we deliberately aligned the long straight edges of top (bottom) graphene sheets to those of the top (bottom) h-BN layers in assembly process[7]. We defined the geometry of devices (Supplementary Fig. S4) using dry etching with $CHF_3/O_2$ in an inductively coupled plasma system after standard electron beam lithography, followed by the deposition of edge contacts via standard electron beam evaporation of Cr (5nm)/Pd (15nm)/Au (30nm).

**Transport measurements**

We performed four-terminal transport measurements at various temperature in an Oxford cryostat, using the low-frequency (17.7Hz) lock-in technique. We applied a current excitation of 4nA in all resistance measurements. We measured the current and voltage using current (voltage) preamplifiers (SR570/560) combined with dual-channel lock-in amplifiers (SR830/865). The measurements of current and voltage were synchronized by a common external TTL signal. We read the results directly from the microprocessors of the lock-in amplifiers to suppress noise during signal conversion. The dual-gate voltages were applied using the Keithley 2636B source measure units. We controlled the relative angle between our device and the magnetic field by mounting the device on a dual axis rotator probe.

The dual-gate structure allows to tune the external carrier density $n_{ext}$ and the external displacement field $D_{ext}$ independently according to the relations $en_{ext} = c_T V_{TG} + c_B V_{BG}$, $D_{ext} = (c_T V_{TG} - c_B V_{BG})/2\varepsilon_0$. Here, $\varepsilon_0$ is the vacuum permittivity, and $c_T, c_B$ are the geometric capacitance per unit area of the top and bottom gates calculated from the thickness of corresponding h-BN crystals.

We present dual-gate mappings obtained in two mapping modes — differed by the sweeping speed of $V_{BG}$. In the fast-sweeping mode, $V_{TG}$ is fixed when $V_{BG}$ sweeps at a speed ~5mV/(nm·s) and $V_{TG}$ only changes by a small amount between two $V_{BG}$ sweeping cycles. Such fast-sweeping mode leads to results shown in Supplementary Fig. S2, where the hysteresis takes the form of a parallelogram on the $V_{TG}$-$V_{BG}$ plane. In the slow-sweeping mode, $V_{BG}$ only changes by a small amount between two $V_{TG}$ sweeping cycles, yielding an average sweeping speed of $V_{BG}$ ~0.3mV/(nm·s). Such slow-sweeping mode leads to results shown in Supplementary Fig. S5, S6 and S7, where the hysteretic features take more of an irregular shape. The difference of hysteresis related to the sweeping speed is also reported in similar moiré systems[3,7], which has been attributed to asymmetric screening of the dual gates. We notice that the unusual magnetoelectric effect is only observed in the slow-sweeping mode.

**Data availability:**
The data that support the plots within this paper and other findings of this study are available from the corresponding authors upon reasonable request.

**Acknowledgments:** This work was supported in part by the National Key R&D Program of China under Grant 2023YFF1203600, the National Natural Science Foundation of China (12322407, 62122036, 62034004, 61921005, 12074176), the Strategic Priority Research Program of the Chinese Academy of Sciences (XDB44000000). F.M. would like to acknowledge support from the AIQ Foundation. We thank Jianpeng Liu for fruitful discussions, and Micro Fabrication Center, Collaborative Innovation Center of Advanced Microstructures for aid in the micro fabrication of the studied devices.

**Author contributions**

F.M., B.C. S.-J.L and M.C. conceived the idea and designed the experiments. F.M., B.C. and S.-J.L. supervised the whole project. Y.X. and M.C. fabricated device. M.C. performed the measurements. F.C., Q.L. and J.X. provided assistance in the experiments. T.T. and K.W. provided h-BN samples. M.C., B.C., S.-J.L. and F.M. co-wrote the manuscript.

**Competing interests**

The authors declare no competing interests.

**Additional information**

**Correspondence and requests for materials** should be addressed to Bin Cheng, Shi-Jun Liang or Feng Miao.